\documentclass[]{aipproc}
\layoutstyle{6x9}

\usepackage{times,colordvi,amsmath,epsfig,float,color,multicol,subfigure,natbib}
\usepackage[latin1]{inputenc}

\newcommand{\apj}{{\it ApJ}}
\newcommand{\gcn}{{\it GCN}}

\def\chandra{{\it Chandra}}
\def\xmm{{\it XMM}}
\def\lessim{\mathrel{\hbox{\rlap{\hbox{\lower4pt\hbox{$\sim$}}}\hbox{$<$}}}}
\def\gtrsim{\mathrel{\hbox{\rlap{\hbox{\lower4pt\hbox{$\sim$}}}\hbox{$>$}}}}

\begin{document}

\title
      [X-ray Lines]
      {Previously Claimed(/Unclaimed) X-ray Emission Lines in High Resolution Afterglow Spectra}

\author{N. Butler}{
  address={Center for Space Research, Massachusetts Institute of Technology, MA},
}

\iftrue
\author{A. Dullighan}{
  address={Center for Space Research, Massachusetts Institute of Technology, MA},
}
\author{P. Ford}{
  address={Center for Space Research, Massachusetts Institute of Technology, MA},
}
\author{G. Ricker}{
  address={Center for Space Research, Massachusetts Institute of Technology, MA},
}
\author{R. Vanderspek}{
  address={Center for Space Research, Massachusetts Institute of Technology, MA},
}

\author{K. Hurley}{
  address={Space Sciences Laboratory, Berkeley, CA},
}
\author{J. Jernigan}{
  address={Space Sciences Laboratory, Berkeley, CA},
}

\author{D. Lamb}{
  address={Department of Astronomy, University of Chicago, IL},
}

\fi

\copyrightyear  {2003}

\begin{abstract}
We review the significance determination for emission lines in
the \chandra~HETGS spectrum for GRB~020813,
and we report on a search for additional lines in high resolution
Chandra spectra.  No previously unclaimed features are found.
We also discuss the significance of lines sets reportedly discovered 
using XMM data for GRB~011211 and GRB~030227.  We find that these 
features are likely of modest, though not negligible, 
significance.  

\end{abstract}

\maketitle

\vspace{-0.5in}

\section{I. Introduction}
\label{section:intro_otherspectra}

\vspace{-0.1in}

Multiple luminous X-ray lines have been claimed in spectra taken
with \xmm~of 2 GRB afterglows.  
(GRB~011211, \citet{reeves02}; GRB~030227, \citet{watson03})
The statistical significance of the GRB~011211 lines has been called into
question by \citet{rutledge03}, and we address this question
in Section IV.  In \citet{butler03a}, we discuss high resolution spectra from
\chandra~for the X-ray afterglows to GRB~020813 and GRB~021004, and we
report the discovery of moderately low significance spectral lines 
in the case of GRB~020813.  Our detection (with an 
independent instrument) supports the claimed multiple$-\alpha$ line 
detections in \xmm~data,
and the high spectral resolution facilitates a clearer determination of the
line significances (see Section II).  There are
3 additional bursts with high resolution \chandra~spectra (GRB~991216,
GRB~020405, and GRB~030328), which we have
analyzed to search for discrete spectral features.  
Only the data for GRB~991216 yielded a claimed line 
detection in the literature \citep{piro00}.  My collaborators and I 
analyzed the GRB~030328 data \citep{butler03b,ford03,butler03c} and 
reported no line detections.  Null detections for GRB~020405 are 
reported by \citet{mirabel02}.

\vspace{-0.3in}

\section{II. S and Si lines for GRB~020813?}
\label{section:020813}

\vspace{-0.1in}

In \citet{butler03a}, we describe our data reduction and continuum fits
for the \chandra~HETGS observation of GRB~020813 (and GRB~021004).  
We also describe our emission line search method which involves successively
binning the spectral data by factors of 2 in order uncover deviations
from the continuum fit.  This procedure is sensitive to resolved emission
lines.
It allows for quick and easy determination of line significances, 
easily verified by Monte Carlo.  One prominent emission line was found for
GRB~020813.  Considering the number of wavelength bins
searched, we estimate a 
multiple-trial (i.e. blind search) significance 
for the line of
$3.3\sigma$.  If the line is identified with the K$\alpha$  transition
in H-like S, a low significance line possibly due to the K$\alpha$ transition
in H-like Si can be identified.  The significance of the pair turns out
to be modestly better ($3.5\sigma$) than the significance of the S line 
alone \citep{butler03a}.

Our S line significance determination agrees with the estimate made using 
the deprecated likelihood ratio test \citep[see, e.g.,][]{protassov02}.
To check whether this fact is statistically meaningful,
we apply Monte Carlo integration to establish the true distribution for the
log-likelihood.  We form $10^4$ simulated data sets
using power-law model parameters (with Galactic absorption) drawn from the 
posteriori distribution (the distribution of model 
parameters given the observed data).  Each data set is then fitted with 
this model, then with this model plus a Gaussian emission line. The
number of iterations yielding larger improvements in $\Delta\chi^2$ than
the observed value ($\Delta\chi^2$ =15.5 for 3 additional degrees of freedom)
is recorded.  To ensure that the parameter space for
the emission line is adequately explored in each Monte Carlo iteration,
we use FFTs to determine the most significant line-like residual on a
fine line centroid wavelength grid ($\delta\lambda=0.01$\AA) and line 
width grid (dyadic
intervals from $\sigma_{\lambda}=0.02$\AA~to $\sigma_{\lambda}=10.24$\AA).  
We find that 13 of $10^4$ runs
yield a larger $\Delta\chi^2$ than the observed value, consistent with the 
significance estimates
quoted above.  We find consistent results independent of whether uniform or
delta function priors are assumed on the power-law model parameters, 
indicating that the parameters are well constrained by the data.

\vspace{-0.3in}

\section{III. Lines in Other High Resolution Chandra Spectra}
\label{section:others}

\vspace{-0.1in}

\begin{table}[t]
\centering
\begin{tabular}{rcccc}
\hline
Afterglow Source & Lines Detected & $z_{\rm x-ray}$ & $z_{\rm optical}$ & Significance \\
\hline
GRB~991216 & Fe (I--XXVI) & $\sim$1 & $\ge 1.02$ & $>3.7\sigma$ \\
GRB~020405 & Ar XVIII, Mg XI, Mg XII & $0.63\pm 0.03$ & 0.695 & $<2\sigma$ \\
GRB~030328 & Mg XI & 1.52 & $\ge 1.52$ & $<1\sigma$ \\
\hline
\end{tabular}
\caption{
\small
Lines detected in the spectra of GRB~991216, GRB~020405,
and GRB~030328 are shown along with the implied redshift ($z_{\rm x-ray}$).
This $z$ can be compared to that measured in the optical ($z_{\rm optical}$).
For GRB~991216, the data do not constrain the ionization state of the
candidate Fe line.
}
\label{table:3grbs_results}
\end{table}

We apply the methods which led to the discovery of the lines for 
GRB~020813 to the 4 other high resolution spectra
taken with \chandra~of GRB X-ray afterglows (GRB~991216, GRB~020405,
GRB~021004, and GRB~030328).
We reduce the data as described in \citet{butler03a}, with
the exception that we jointly fit the gratings data along with the 0th-order 
data, instead of fitting the gratings data alone, in order to best determine 
the continuum fits.
The continuum fits and the line searches in the gratings data
are described in detail in Nat Butler's Ph.D. thesis \citep{natthesis}.  The
results are shown in Table \ref{table:3grbs_results}.  No highly significant,
previously
unreported lines are detected, and the Fe line reported by \citet{piro00}
is robustly detected with a multiple trials significance (conservatively)
better than $3.7\sigma$.  This line is apparent in each of the 4
independent 1st-order HETGS spectra, and it is also seen in the 0th-order
data.  It is, in our opinion, the best case for an emission line in 
a GRB X-ray afterglow to date.

\vspace{-0.3in}

\section{IV. The Significance of the XMM Multiple$-\alpha$ Lines}
\label{section:2xmmlines}

\vspace{-0.1in}

We reduce the EPIC-pn data for GRB~011211 following \citet{reeves02b},
finding 537 net counts in the first 5 ksec of the observation.
To search for emission lines, we employ the ``matched filter'' 
technique described in \citet{rutledge03}.  We correct a number of
minor errors in that work: (1)
\citet{rutledge03} base their significance estimates on model continua
determined using $\chi^2$ fits of 
sparsely binned ($\sim 12$ counts/bin) data.
We find that the grouped data bin boundaries are not robust and
that modest shifting can arise 
due to minor changes in the source and background selection regions.
This shifting is sufficient to occasionally wash out 
the indication of line emission.  We choose to fit continuum models to
the unbinned spectrum.
(2)
\citet{rutledge03} uniformly sample from a number of possible continuum
models rather than sampling according to the
posteriori distribution of possible models given the observed
data. For example, in the case of $\chi^2$ fitting, our approach would
suggest sampling from a model A 
in frequency relative to a model B as 
$\exp{\left \{-0.5\chi_A^2/\chi_B^2\right \}}$, assuming uniform priors
on the model parameters.
(3)
\citet{rutledge03} also unjustifiably increase the normalizations on their 
continuum models (relative to the best fit values) in order to force each 
to yield the total number of observed source counts.  

\begin{figure}[ht]
\centering
\rotatebox{0}{\resizebox{18pc}{!}{\includegraphics{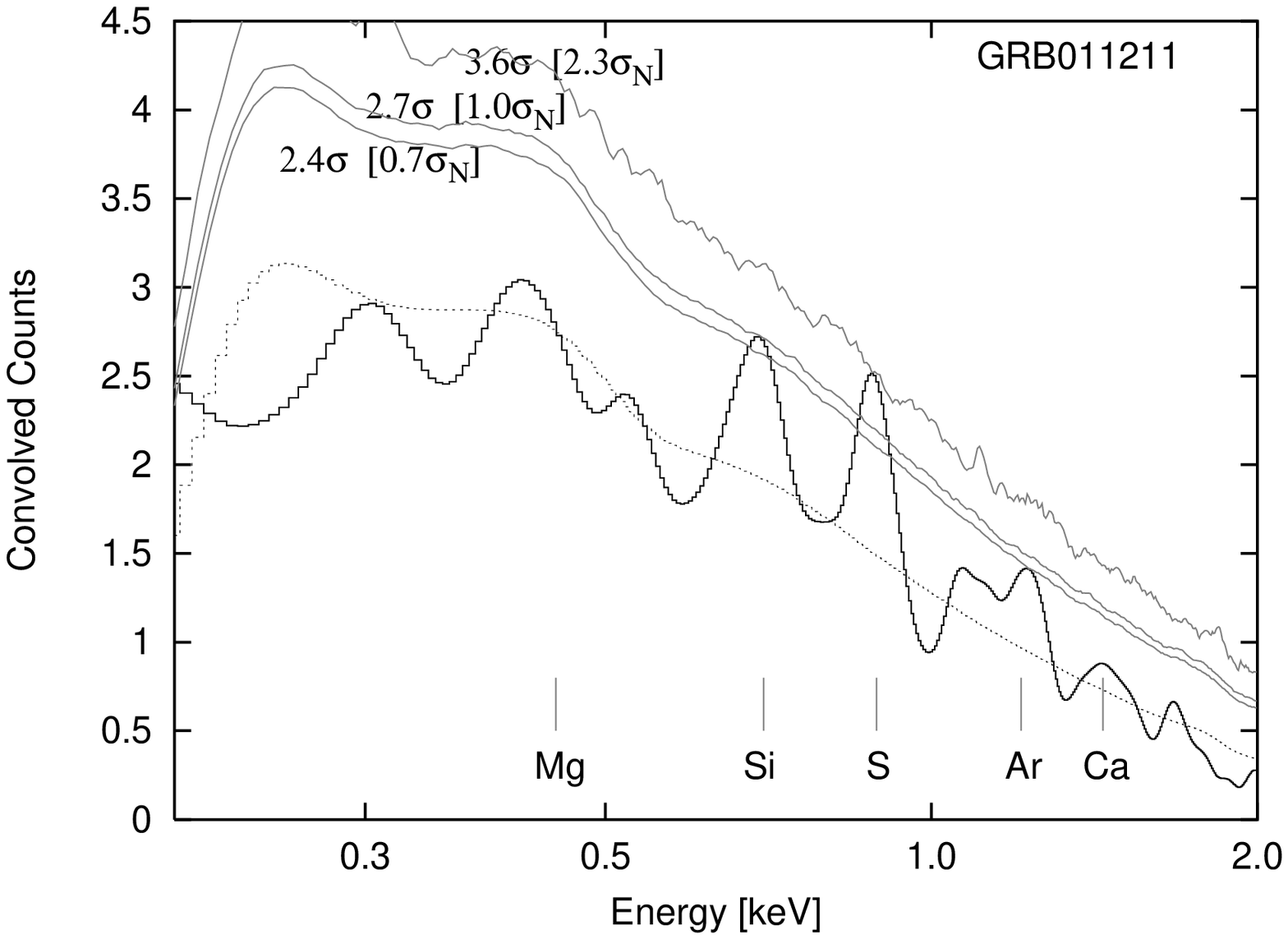}}}
\rotatebox{0}{\resizebox{18pc}{!}{\includegraphics{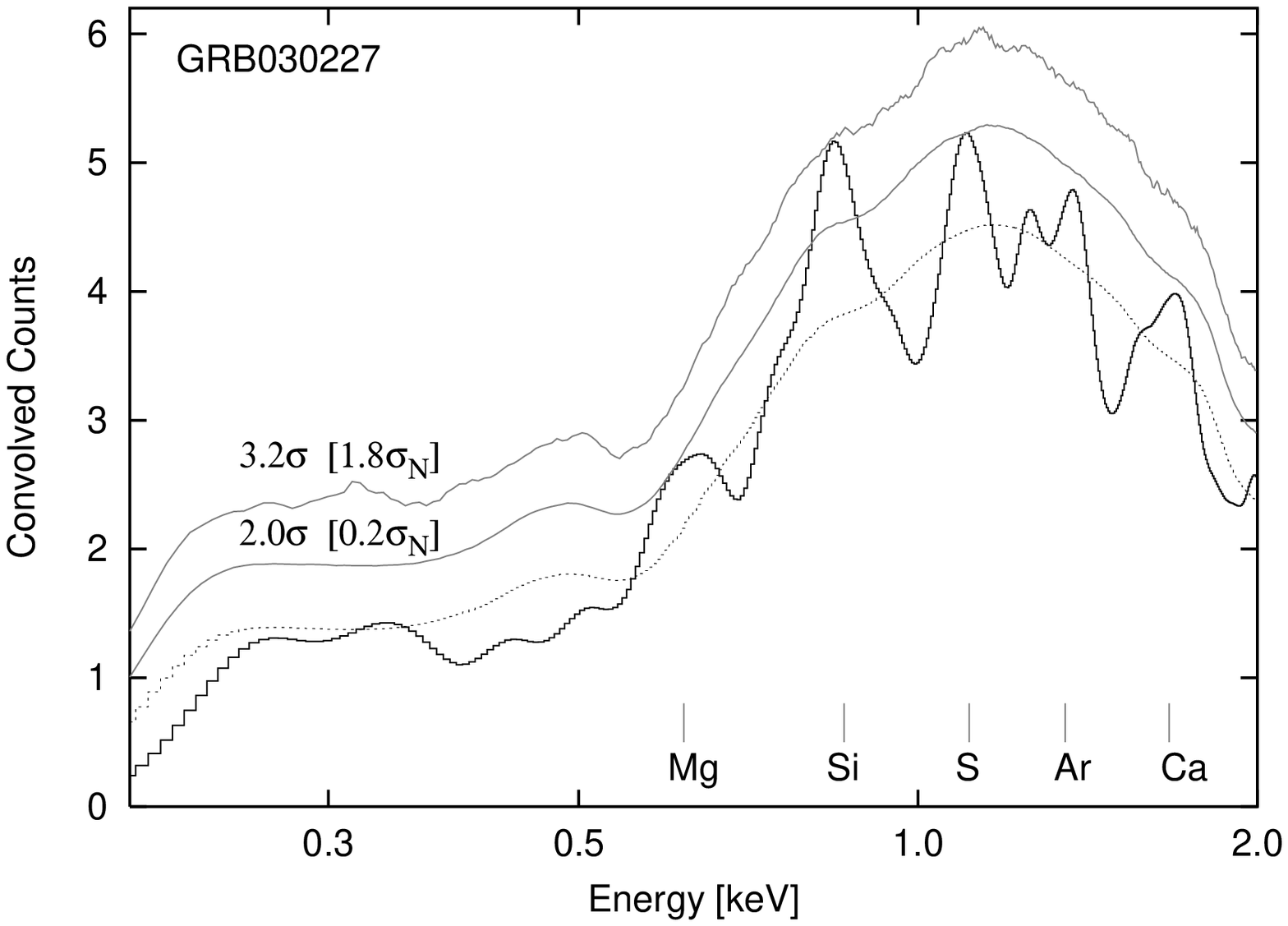}}}
\caption{
\small
The raw PI spectra for 2 \xmm~X-ray afterglows are convolved with
the line response function (i.e. ``the matched filter'' of
\citet{rutledge03}, dark curve).
Significance contours are calculated using an MCMC method, which does
not rely on binning the data.  The locations of emission lines claimed
by \citet{reeves02} and \citet{watson03} are indicated.
}
\label{fig:xmm_lines}
\end{figure}

We model the continuum emission as an absorbed power-law, with the
absorption fixed at the Galactic value as in \citet{reeves02b}.
We apply the
Markov Chain Monte Carlo (MCMC) algorithm described in Appendix B of 
\citet{vandyke01} to determine the distribution of null model parameters
given the data in raw PI bins.  The background is modelled using a 
broken power-law with
a break at 1.35 keV as in \citet{rutledge03}.  We consider a uniform
(i.e. totally non-informative) prior distribution on each model parameter 
for the source and background counts.  From $10^4$ models 
generated in the fashion, we simulate $10^4$ spectra and 
apply the matched filter of \citet{rutledge03} to each.

The first panel in Figure \ref{fig:xmm_lines}
shows the result of applying this filter to the observed data 
(dark line).  The mean result from the simulated spectra is plotted as
a dotted line. 
At the locations of the reported S, Si, and Ar emission lines
\citep[see,][]{reeves02}, the solid dark curve in the first panel of
Figure \ref{fig:xmm_lines} deviates from the mean at $>2\sigma$ significance.
The significance contour reached by each feature is plotted, and
the individual (single-trial) 
significances of the lines are $3.6\sigma$, $2.7\sigma$, and $2.4\sigma$.
Generating an additional $10^4$ model sample parameters and reapplying 
the matched filter to the simulated spectra for each model, we count the
number of times that a simulated model breaks through each significance
contour between 0.2 and 2.0 keV.   This allows
us to determine that the multiple trials significances ($\sigma_N$ values
in Figure \ref{fig:xmm_lines}) for the emission
lines are $2.3\sigma_N$, $1.0\sigma_N$, and $0.7\sigma_N$.
The significance of the line set
can then be estimated as 5 times the product of these probabilities
(i.e. $2.4\sigma$), where the factor of 5 takes into account that the
data have been analyzed in 5 time regions.  This is higher than
the $\sim 1.0\sigma$ value determined by \citet{rutledge03}, and
it is lower than the $3.9\sigma$ initially suggested by \citet{reeves02}.  

Recently, \citet{watson03} have claimed the detection of a line set
in GRB~030227, remarkably similar to the detections claimed for GRB~011211 by
\citet{reeves02}.  \citet{watson03} quote a significance determined
solely via the likelihood ratio test ($2.7-4.4\sigma$, depending
on the number of degrees of freedom).  
We reduce the EPIC-pn data for GRB~030227 following \citet{watson03},
finding 1593 source counts in the final 10 ksec of the observation.
In our MCMC analysis, we describe the background counts as a power-law.
The source counts are modelled as an absorbed power-law as in \citet{watson03}.
All model parameter prior distributions are taken as uniform.
As displayed in the second panel of Figure \ref{fig:xmm_lines}, 
the claimed Mg, Si, and S lines have significances $\gtrsim 2.0\sigma$
(single-trial).  This data set is the last of 4 time slices within the 
observation.  Performing the multi-trial significance calculation as
above, we find the the significance of the line set is $1.3 \sigma$.
Thus, we find that the 
line emission for GRB~030227 is less significant than that for GRB~011211.
We stress that this significance estimate, like that made for GRB~011211
above, is a lower bound on the significance of the line emission.  If
the line centroids can be argued to be constrained by the physics,
as argued in \citet{watson03}, then the significance would increase.

\vspace{-0.3in}

\section{V. Conclusions}

\vspace{-0.1in}

We conservatively estimate the significances for line emission 
in GRB~991216 (\chandra), GRB~020813 (\chandra), GRB~011211 (\xmm), 
and GRB~030227 (\xmm) as $>3.7\sigma$, $3.5\sigma$, $2.4\sigma$, and 
$1.3\sigma$, respectively.  
We do not find the line emission for GRB~011211 (which has galvanized
many GRB researchers) to be entirely insignificant, as claimed by
\citet{rutledge03}.  However, we find none of the 
observations to be highly significant.
Hopefully, early observations with Swift in the coming few years will decide
conclusively whether this emission is real.

\end{document}